\pdfoutput=1
\documentclass[aps,pra,reprint,a4paper,superscriptaddress,floatfix,amsmath,amssymb,amsfonts,noshowpacs,]{revtex4-1}
\usepackage{newtxtext,newtxmath}
\usepackage[utf8]{inputenx}
\usepackage{graphicx}
\usepackage[dvipsnames]{xcolor}
\usepackage{soul} 
\usepackage[textwidth=17.5cm,textheight=23.5cm,verbose,pdftex]{geometry}
\usepackage{floatrow}
\usepackage[pdftex]{hyperref}

\hypersetup{pdfauthor={Felix Feix}, bookmarksnumbered=true, pdftitle={Ga-polar (In,Ga)N/GaN quantum wells vs.\ N-polar (In,Ga)N quantum disks in GaN nanowires: Comparative analysis of carrier recombination, diffusion, and radiative efficiency}, colorlinks, citecolor=blue, linkcolor=blue, urlcolor=blue}

\begin{document}
\title{Ga-polar (In,Ga)N/GaN quantum wells vs.\ N-polar (In,Ga)N quantum disks in GaN nanowires: Comparative analysis of carrier recombination, diffusion, and radiative efficiency}
\author{F. Feix} \email[Electronic mail: ]{feix@pdi-berlin.de}
\author{T. Flissikowski}
\affiliation{Paul-Drude-Institut für Festkörperelektronik, Leibniz-Institut im Forschungsverbund Berlin e.\,V., Hausvogteiplatz 5--7, 10117 Berlin, Germany}
\author{K. K. Sabelfeld}
\affiliation{Institute of Computational Mathematics and Mathematical Geophysics, Russian Academy of Sciences, Novosibirsk, Russian Federation}
\author{V. M. Kaganer}
\author{M. Wölz}
\altaffiliation{Present address: Jenoptik Laser GmbH, Göschwitzer Str.\ 29, 07745 Jena, Germany}
\author{L. Geelhaar}
\author{H. T. Grahn}
\author{O. Brandt}
\affiliation{Paul-Drude-Institut für Festkörperelektronik, Leibniz-Institut im Forschungsverbund Berlin e.\,V., Hausvogteiplatz 5--7, 10117 Berlin, Germany}


\begin{abstract}
We investigate the radiative and nonradiative recombination processes in planar (In,Ga)N/GaN(0001) quantum wells and (In,Ga)N quantum disks embedded in GaN$(000\bar{1})$ nanowires using photoluminescence spectroscopy under both continuous-wave and pulsed excitation. The photoluminescence intensities of these two samples quench only slightly between 10 and 300~K, which is commonly taken as evidence for high internal quantum efficiencies. However, a side-by-side comparison shows that the absolute intensity of the Ga-polar quantum wells is two orders of magnitude higher than that of the N-polar quantum disks. A similar difference is observed for the initial decay time of photoluminescence transients obtained by time-resolved measurements, indicating the presence of a highly efficient nonradiative decay channel for the quantum disks. In apparent contradiction to this conjecture, the decay of both samples is observed to slow down dramatically after the initial rapid decay. Independent of temperature, the transients approach a power law for longer decay times, reflecting that recombination occurs between individual electrons and holes with varying spatial separation. Employing a coupled system of stochastic integro-differential equations taking into account both radiative and nonradiative Shockley-Read-Hall recombination of spatially separate electrons and holes as well as their diffusion, we obtain simulated transients matching the experimentally obtained ones. The results reveal that even dominant nonradiative recombination conserves the power law decay for (In,Ga)N/GaN\{0001\} quantum wells and disks.
\end{abstract}

\maketitle

\section{INTRODUCTION}\label{sec:introduction}
The active region in devices for solid-state lighting \cite{Nakamura1994,Narukawa2010}, display technologies \cite{Pust2015}, and diode lasers \cite{Nakamura1996} is formed by quantum wells (QWs) consisting of the ternary alloy (In,Ga)N. A record-high external quantum efficiency (EQE) of about 84\% for a blue (In,Ga)N-based light emitting diode (LED) has been achieved already in 2010 \cite{Narukawa2010}. However, the realization of phosphor-free white LEDs with both a high luminous efficiency and a high color rendering index requires the use of efficient narrow-band emitters not only in the blue, but also in the green and red spectral range \cite{Taylor2012,Ra2016}. For this reason, the U.S. Department of Energy has released a research and development plan that amongst others prioritizes the development of efficient green emitter materials \cite{Bardsley2015} to overcome the so-called ``green gap'', which denotes the drastic reduction of the luminous efficiency of (In,Ga)N as well as (Al,In,Ga)P LEDs in the green spectral range \cite{Bulashevich2015,Weisbuch2015}. For both materials, this phenomenon is caused by a steep decline of the internal quantum efficiency (IQE) for wavelengths approaching the green spectral range. In (In,Ga)N-based LEDs, the potential reasons for this decline with increasing In content are manifold and include a possible deterioration of the crystal quality resulting in an increase of defect-assisted nonradiative processes \cite{Singh1997} as well as a reduced radiative rate due to an increasing magnitude of the polarization fields \cite{Takashi1998} and localization phenomena \cite{AufderMaur2016}.

Since both of these issues are directly related to the increasing magnitude of strain in the (In,Ga)N layer, axial (In,Ga)N/GaN$(000\bar{1})$ nanowire heterostructures are considered as a promising alternative to planar structures for long-wavelength emission \cite{Jahangir2014}. In molecular beam epitaxy (MBE), N-polar GaN nanowires spontaneously form on various technologically  attractive substrates such as Si \cite{*[{See:  }] [{ and references therein. }] Romanyuk2015} or metal foils \cite{May2016,Calabrese2016} while retaining their single-crystal nature. Using MBE, (In,Ga)N quantum disks (QDs) can be subsequently synthesized on the GaN nanowires. In sufficiently thin nanowires, the lattice mismatch between the (In,Ga)N QD and GaN is partly accommodated elastically due to strain relaxation at the nanowire sidewalls \cite{*[{See:  }] [{ and references therein. }] Krause2016}. This strain relief facilitates the incorporation of high In contents without the formation of extended defects, reduces the driving force to generate point defects, and decreases the magnitude of the polarization field in the (In,Ga)N QD embedded within the nanowire. Moreover, nanowires naturally exhibit a much higher extraction efficiency compared to planar samples \cite{*[{See:  }] [{ and references therein. }] Hauswald2017}. Indeed, several groups have reported nanowire LEDs on Si emitting in the green and even red spectral range with IQEs up to 50\% \cite{Guo2011,Nguyen2012,Jahangir2014,Sarwar2015}. At the same time, the EQE of these devices has so far remained significantly lower than that of conventional planar LEDs at comparable wavelengths. For example, yellow LEDs were reported in Ref.~\cite{Janjua2016} with values for the IQE and EQE of 40\% and 0.014\%, respectively. It is crucial to elucidate the origin of this blatant discrepancy for a realistic assessment of the potential of axial nanowire heterostructures for efficient full-color emitters.

In this paper, we focus on the investigation of the radiative and nonradiative recombination processes for a representative spontaneously formed GaN$(000\bar{1})$ nanowire ensemble containing (In,Ga)N QDs by temperature-dependent photoluminescence (PL) spectroscopy under both continuous-wave (cw) and pulsed excitation. All measurements are performed in the form of a side-by-side comparison with a state-of-the-art planar (In,Ga)N/GaN(0001) QW structure to allow a determination of the actual absolute luminous intensity of the nanowire sample. In addition to an estimate of the IQE of the nanowire sample, our experiments provide important insight into the role of carrier localization and diffusion on the recombination dynamics in both the planar Ga-polar QWs and the N-polar QDs embedded in GaN nanowires.

The paper is organized as follows. Section~\ref{sec:methods} provides details regarding the methods used in our work. Temperature- and excitation-dependent cw-PL spectra and integrated intensities of both samples are discussed in Sec.~\ref{sec:basic}. Section~\ref{sec:trpl-ini} is devoted to the analysis of the initial (ns) decay in PL transients obtained under pulsed excitation. Transients over the full time range (\textmu s) are presented in Sec.~\ref{sec:powerlaw}. The peculiar power law decay observed in these transients for both samples is modeled by diffusion-reaction equations taking into account radiative and nonradiative recombination as well as carrier diffusion. Section~\ref{sec:summary} summarizes our results and draws conclusions from the results of the simulation of the experimental PL transients.

\section{SAMPLES, EXPERIMENTS, AND METHODS} \label{sec:methods}
As a reference, we use a planar (In,Ga)N/GaN(0001) quantum well structure with an In content of about 0.15 which was grown by metal-organic chemical vapor deposition (MOCVD) on a Si(111) substrate. The sample was fabricated in an LED production reactor and has an IQE close to the state of the art for blue emitting LEDs in 2012 \cite{Schiavon2013}. The axial (In,Ga)N/GaN(000$\bar{1}$) nanowire heterostructure we have chosen for this study is representative for this type of samples and was grown by MBE on a Si(111) substrate. The In content in the QDs was determined by x-ray diffractometry to amout to $0.26 \pm 0.1$. The sample was selected by virtue of its comparatively high luminous efficiency and the fact that is has been very thoroughly investigated by both structural and spectroscopic techniques \cite{Laehnemann2014}.

Both samples contain similar active regions, consisting of five (six) 3-nm-thick (In,Ga)N QWs (QDs) separated by undoped GaN barriers with thicknesses larger than 7\,nm for the planar (nanowire) structure. The nanowire ensemble has a surface coverage of around 50\%. Hence, the active volume of both samples differs by less than a factor of two. X-ray diffractometry and transmission electron microscopy performed on these types of samples demonstrate that the (In,Ga)N/GaN interfaces are coherent and exhibit no misfit dislocations. The density of threading dislocations is in the $10^8$\,cm$^{-2}$ range for the planar sample and is basically zero within the top part (containing the QDs) of the nanowire ensemble.

In contrast to planar (In,Ga)N/GaN QWs that reach their maximum performance in the blue spectral range \cite{Schiavon2013}, axial insertions in nanowires show a higher luminous efficiency in the green because of a complex interplay of surface potentials and polarization fields \cite{Marquardt2013}. We have selected the two samples used in this study accordingly: both belong to the brightest emitters for their class. The two samples were measured mounted side by side, and the PL signal was corrected for the spectral response of the detection system to obtain a meaningful comparison. The measured PL intensity also depends on the absorbance of the structure at the wavelength of the laser used for excitation as well on the efficiency with which the internally emitted radiation is extracted. Due to light scattering and diffraction, both of these quantities are enhanced for nanowires, particularly so the extraction efficiency \cite{Pfueller2012,Hauswald2017}. For simplicity, however, we assume in all what follows that light absorption and extraction are comparable for the two samples under investigation. Thus, when comparing external luminous efficiencies, we are overestimating the IQE of the nanowire sample.

For all PL experiments, the samples were mounted in a He-flow cryostat allowing a continuous variation of temperature between 10 and 300\,K. Cw-PL spectroscopy was performed utilizing quasi-resonant excitation of the samples by a diode laser ($\lambdaup_L = 402$\,nm) with excitation power densities ranging from Wcm$^{-2}$ to MWcm$^{-2}$. We also performed experiments with nonresonant excitation ($\lambdaup_L = 325$\,nm) and obtained essentially identical results (not shown in this paper). The laser was focused and the cw-PL signal was collected by a microscope objective in a confocal arrangement. The signal was dispersed by a monochromator with a spectral resolution of about 0.4\,meV and detected with a liquid-nitrogen-cooled charge-coupled device.

For time-resolved PL experiments, a frequency-doubled, femtosecond Ti:sapphire laser ($\lambdaup_L = 349$\,nm) was employed. The PL was focused and detected with a microscope objective. The transients were recorded at the respective peak energies integrated over a spectral range of about 20\,meV \footnote{Transients recorded at the low- or high-energy tails of the PL band (that do not  substantially contribute to the integrated PL intensity) will change the shape of the transient slightly, as we probe localization centers with different confining potential.}. The energy fluence per pulse amounted to approximately 3\,\textmu J\,cm$^{-2}$, corresponding to a maximum charge carrier density of approximately $3\times 10^{11}\,$cm$^{-2}$. The excitation density was chosen such that no spectral diffusion of the PL band was detected, i.\,e., a screening of the internal electric fields and hence a dynamically changing overlap of the electron and hole wave functions was avoided. In addition, the low repetition rate (9.3\,kHz for recording the transients up to 33\,\textmu s) of the laser pulses prevented the accumulation of charge carriers created by consecutive laser pulses. By employing time-correlated single photon counting in conjunction with an appropriate binning in the time-domain and a careful subtraction of the background, we achieve a dynamic range in the detection of six orders of magnitude. The temporal resolution amounted to 45\,ps for the investigation of the initial decay of the PL and was decreased to 400\,ps when recording transients over the full time range of 33\,\textmu s.

For the simulation of the decay kinetics, a coupled system of stochastic integro-differential equations was solved by the Monte Carlo algorithm developed by \citet{Sabelfeld2015}. The Fortran 95 code was executed on a 4-way Intel\textsuperscript{\textregistered} Xeon\textsuperscript{\textregistered} E5-4627v2, allowing us to track the radiative and nonradiative annihilation of 400 randomly situated electrons and holes as well as their diffusion. For each parameter set, the computation was repeated 1000 times with a random seed to obtain sufficient statistics. The complete simulation of a transient typically required about 10~min on a single core of the system.

\begin{figure*}
\floatbox[{\capbeside\thisfloatsetup{capbesideposition={right,center},capbesidewidth=4.5cm}}]{figure}[\FBwidth]
{\caption{(a) Cw-PL spectra (over energy $E$ and wavelength $\lambda$) of the planar and the nanowire sample excited with an intensity of approximately 50\,W\,cm$^{-2}$ at a temperature $T$ of 10 and 300\,K in a semi-logarithmic representation. (b) Normalized Arrhenius representation of the integrated PL intensity $I_\text{PL}$ (symbols) measured with an excitation density of approximately 100\,W\,cm$^{-2}$. Solid lines show fits to the data as discussed in the text. The extracted activation energies amount to 48 and 34\,meV for the planar and the nanowire sample, respectively. (c) $I_\text{PL}$ vs. the excitation intensity $I_\text{exc}$ acquired at 300\,K. The solid lines indicate fits whose slopes are indicated in the figure.}
\label{fig:cw-pl}}
{\includegraphics[width=12.2cm]{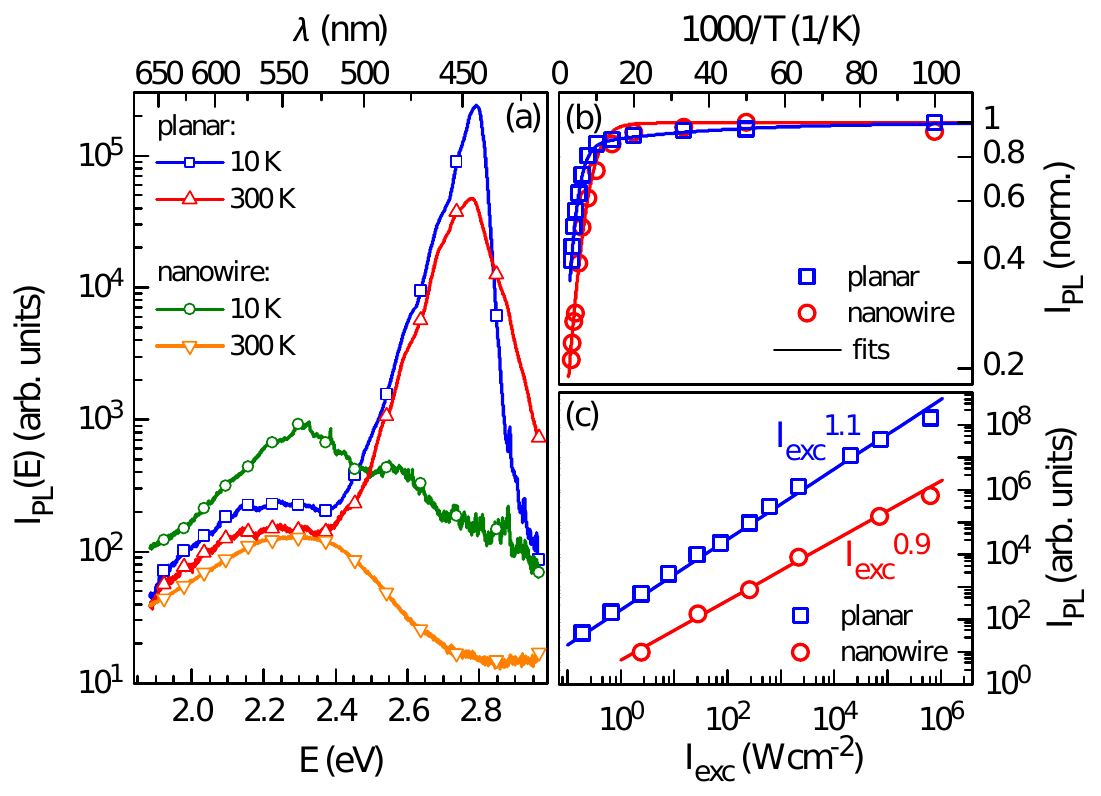}}
\end{figure*}

\section{RESULTS AND DISCUSSION}\label{sec:results}

\subsection{Basic cw-PL characterization}\label{sec:basic}
Figure \hyperref[fig:cw-pl]{\ref*{fig:cw-pl}(a)} shows exemplary cw-PL spectra of the planar and the nanowire sample on a semi-logarithmic scale. The most obvious and striking difference between the two samples is the PL intensity, which is two orders of magnitude higher for the planar sample at both low (10\,K) and high (300\,K) temperatures. The low-temperature PL band of the QWs of the planar sample peaks at 2.795\,eV with a full width at half maximum (FWHM) of 70\,meV. Compared to the binary compound GaN, the PL band is broad reflecting the inherent alloy disorder in the ternary compound (In,Ga)N \cite{Zeng1998}. The band redshifts with increasing temperature by about 20\,meV, 
broadens to 130\,meV, and decreases in intensity. The broad luminescence band at 2.2\,eV is caused by yellow luminescence in the GaN buffer layer as commonly observed in MOCVD-grown GaN \cite{Reshchikov2005a}. The dominant, broad PL band of the MBE-grown nanowires peaks at 2.32\,eV and exhibits an FWHM of 300\,meV at 10\,K. This large line width is not only caused by alloy disorder, but also by variations in the In content and the quantum disk width between the individual nanowires \cite{Laehnemann2011}. Due to the small number of nanowires probed in these cw-PL experiments with a \textmu m-sized excitation spot, individual spikes due to localized states, particularly at the high energy part of the spectrum, can be observed at low temperatures and low excitation densities. For even lower excitation densities, these spikes can dominate the spectrum entirely as found in a previous study of the same sample \cite{Laehnemann2014}. With increasing temperature, the band redshifts by 30\,meV and decreases in intensity similarly to the planar reference sample.

The decrease in intensity observed for both samples reflects the presence of nonradiative recombination channels at elevated temperature. In fact, the analysis of the temperature-dependent PL intensity is a frequently employed method to study the impact of nonradiative recombination in semiconductors. Nonradiative recombination is often assumed to be thermally activated, and to be negligible at low temperatures \cite{Hangleiter2004}. The latter assumption is not based on any sound physical argument, but if we accept it for the moment, a quenching of the PL intensity at elevated temperatures is directly related to the IQE of the investigated sample. In Fig.~\hyperref[fig:cw-pl]{\ref*{fig:cw-pl}(b)}, we show the temperature-dependent evolution of the normalized integrated PL intensity $I_\text{PL}$ of the planar and the nanowire sample. For both samples, a moderate thermal quenching of the PL intensity is observed. The activation energies, deduced from fits employing the common three-level model of a thermally activated PL quenching \cite{Bimberg1971} [cf.\ Fig.~\hyperref[fig:cw-pl]{\ref*{fig:cw-pl}(b)}], are similar and amount to 48 and 34\,meV for the planar and the nanowire sample, respectively.

If we proceed as it is commonly done and take the ratio of $I_\text{PL}(300\,\text{K})/I_\text{PL}(10\,\text{K})$ as the IQE at room temperature (thus implicitly assuming an IQE of unity at 10\,K \cite{Janjua2016,Hangleiter2004,Guo2010,Hammersley2015,Nguyen2011}), we obtain $\eta = 0.41$ for the planar and $\eta = 0.21$ for the nanowire sample directly from the normalized $I_\text{PL}$ shown in Fig.~\hyperref[fig:cw-pl]{\ref*{fig:cw-pl}(b)}. Obviously, the latter value for the nanowire sample is meaningless considering that the integrated PL intensity of this sample is more than two orders of magnitude lower than that of the planar sample [cf.\ Figs.~\hyperref[fig:cw-pl]{\ref*{fig:cw-pl}(a)} and~\hyperref[fig:cw-pl]{\ref*{fig:cw-pl}(c)}].
This result shows that the ratio $I_\text{PL}(300\,\text{K})/I_\text{PL}(10\,\text{K})$ cannot be taken as a sensible measure for the IQE of samples for which no independent data support the assumption of an IQE of 1 at 10\,K.

\begin{figure}
\includegraphics[width=1\columnwidth]{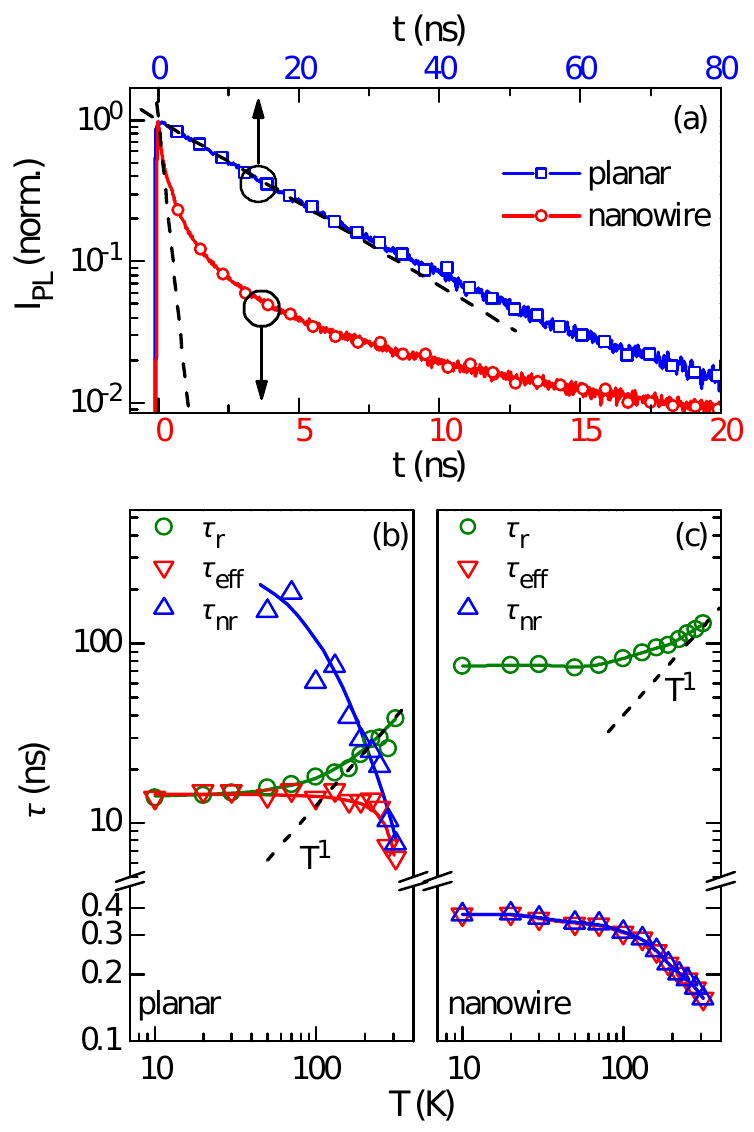}
\caption{(a) Semi-logarithmic representation of the normalized PL transients of the planar and the nanowire sample recorded at 10\,K. Note the different time scales (t) for the planar and the nanowire sample. The dashed lines are fits with single exponentials with different effective lifetimes. (b) and (c) Temperature-dependent effective ($\tau_\text{eff}$), radiative ($\tau_{r}$), and nonradiative ($\tau_{nr}$) PL lifetimes of (b) the planar and (c) the nanowire sample. The solid lines are guides to the eye. The dashed lines illustrate the linear increase of $\tau_{r}$ with increasing temperature $T$. Note the axis break in the y-axis of the double-logarithmic representation.}
\label{fig:pl-lifetimes}
\end{figure}

In Fig.~\hyperref[fig:cw-pl]{\ref*{fig:cw-pl}(c)}, we show the excitation-dependent integrated PL intensity of the planar and the nanowire sample recorded at 300\,K. The spectrally integrated PL intensity $I_\text{PL}$ of the nanowire sample amounts to 1.4\% of the value of the planar sample at low excitation densities, and this ratio decreases to 0.3\% at high excitation densities. Nevertheless, the increase of $I_\text{PL}$ with excitation power density $I_\text{exc}$ is close to linear for both samples over six orders of magnitude ($I_\text{PL} \propto I_\text{exc}^{1.1}$ for the planar and $I_\text{PL} \propto I_\text{exc}^{0.9}$ for the nanowire sample).  We do neither observe a superlinear increase of $I_\text{PL}$ due to a saturation of Shockley-Read-Hall centers \cite{Brandt1996, Reshchikov2013} nor a marked sublinear increase at high excitation densities because of increasing contribution of carrier leakage \cite{Schubert2009} or the onset of Auger recombination \cite{Danhof2011}.


\subsection{Analysis of the initial PL decay}\label{sec:trpl-ini}

The results discussed in Sec.~\ref{sec:basic} show that the temperature and excitation dependence of the cw-PL intensity are not necessarily sensitive to the actual IQE of a given sample. For binary bulk semiconductors such as GaN, for which a unique radiative lifetime exists \cite{Hauswald2014}, a reliable measure of the IQE is the minority carrier lifetime determined by time-resolved PL experiments \cite{Langer2011}. However, in (In,Ga)N/GaN\{0001\} QWs, the radiative lifetime depends strongly on well width and In content due to the presence of large piezoelectric fields. Additionally, the inevitable compositional fluctuations in the random alloy (In,Ga)N lead to the localization of charge carriers \cite{Chichibu2006,Watson-Parris2011,Schulz2015,Schulz2015b}, making the definition of a unique radiative lifetime all but impossible. Still, time-resolved PL measurements in conjunction with a comparative measurement of the absolute emission intensities enable one to disentangle radiative and nonradiative contributions to the PL decay.

In the following, we first focus on the initial part of the PL transients that accounts for a decay in PL intensity of two decades after pulsed excitation. This dynamic range is typical for time-resolved PL experiments for which a streak camera is used for detection as reported, for example, in Refs.~\onlinecite{Langer2014, Hammersley2015, Murotani2015, Pophristic1998, Solowan2013, Aleksiejunas2014}. Figure~\hyperref[fig:pl-lifetimes]{\ref*{fig:pl-lifetimes}(a)} displays PL transients of the planar and the nanowire sample recorded at 10\,K in a semi-logarithmic representation. Evidently, this initial decay of the PL intensity is significantly faster for the nanowire as compared to the planar sample: while it takes 80~ns for the PL intensity of the latter to decrease by two orders of magnitude, it only requires 20~ns for that of the former. Note that the decay cannot be described by a single exponential [see the dashed lines in Fig.~\hyperref[fig:pl-lifetimes]{\ref*{fig:pl-lifetimes}(a)}], particularly so for the nanowire sample. This nonexponential decay is not caused by a screening of the piezoelectric fields, since the transients were acquired with excitation densities well below the onset of this effect.

To derive a decay time despite the nonexponential nature of the transients, we define a phenomenological effective PL lifetime $\tau_\text{eff}$ as the time at which $I_\text{PL}$ has decreased to $1/e$ ($ \approx 37$\%) of its initial value. In addition, we assume that
\begin{align}
\label{eq:1}
\frac{1}{\tau_\text{eff}}=\frac{1}{\tau_r}+\frac{1}{\tau_{nr}}
\end{align}
with the radiative and nonradiative lifetimes $\tau_{r}$ and  $\tau_{nr}$, respectively. Figures \hyperref[fig:pl-lifetimes]{\ref*{fig:pl-lifetimes}(b)} and \hyperref[fig:pl-lifetimes]{\ref*{fig:pl-lifetimes}(c)} show the temperature dependence of $\tau_\text{eff}$ for the planar and the nanowire sample, respectively. For the planar sample, the effective PL lifetime is constant up to 220\,K with a value of about 14\,ns and decreases subsequently to 6\,ns at 300\,K. A similar behavior is observed for the nanowire sample, for which $\tau_\text{eff}$ amounts to about 0.37\,ns between 10 and 70\,K and decreases to 0.16\,ns at 300\,K.

To distinguish the radiative and nonradiative contributions to $\tau_\text{eff}$, we first determine the temperature dependence of the radiative lifetime $\tau_{r}$ from the inverse peak PL intensity of the transient just after the laser pulse \cite{Brandt1996}. To deduce absolute values for $\tau_{r}$ and $\tau_{nr}$, an additional information is required. Since $\tau_\text{eff}$ and $\tau_\text{r}$ are related by
\begin{align}
\label{eq:2}
\eta=\frac{\tau_\text{eff}}{\tau_r},
\end{align}
the required information is the IQE $\eta$ at, for example, 10\,K. This quantity is often indiscriminately assumed to be unity regardless of the sample. In the present case, we use the planar (In,Ga)N/GaN(0001) QWs as reference whose IQE is known to be high even at room temperature: LEDs with these QWs as active region exhibit an EQE between 0.1 and 0.7 at low injection levels \cite{Schiavon2013}. For this sample, it thus seems justified to assign a value of unity to its IQE at 10\,K. To obtain a corresponding value for the (In,Ga)N/GaN$(000\bar{1})$ QDs, we recall that the IQE is proportional to the temporally integrated intensity of the transient. Since this intensity is 200 times lower for the nanowire sample as compared to the planar reference, we thus obtain $\eta \approx 0.005$ for the nanowire sample at 10\,K.

\begin{table}[t]
 \caption{Effective ($\tau_\text{eff}$), radiative ($\tau_{r}$) and nonradiative ($\tau_{nr}$) PL lifetimes extracted for different temperatures $T$ from the analysis of the initial decay of the PL intensity. The IQE $\eta$ at 250\,K is estimated from the initial PL decay.}
 \label{tab:lifetimes_iqe}
\begin{ruledtabular}
  \begin{tabular}{c c c c c c}
sample & $T$ (K) &$\tau_\text{eff}$ (ns) & $\tau_{r}$ (ns) & $\tau_{nr}$ (ns) & $\eta$\\
\hline
planar & 10 & 14 & 14 & $\infty$ & 1 \\
planar & 250 & 12 & 30 & 21 & 0.40 \\
nanowire & 10 & 0.37 & 75 & 0.37 & 0.005 \\
nanowire & 250 & 0.19 & 114 & 0.19 & 0.002 \\
\end{tabular}
\end{ruledtabular}
\end{table}

\begin{figure*}
\floatbox[{\capbeside\thisfloatsetup{capbesideposition={right,center}, capbesidewidth=4.5cm}}]{figure}[\FBwidth]{\caption{Double-logarithmic representation of exemplary experimental PL transients over the full time range of 30\,\textmu s recorded at the PL peak energy of the (a) planar and (b) nanowire sample at 10 and 250\,K. The intensity has been scaled by the same factor to facilitate a direct comparison with the calculations. (c) and (d) Corresponding transient photon fluxes $\phi$ computed by means of Eqs.~(\ref{eq:3})--(\ref{eq:5}). The dotted line in (d) shows a transient for the nanowire sample at 10\,K obtained with values for $b_0$ and $N$ (namely, $b_0'$ and $N'$) that keep the product constant (i.\,e., $b_n' N' = b_n N$).}
\label{fig:exp-simu}}
{\includegraphics[width=11.9cm]{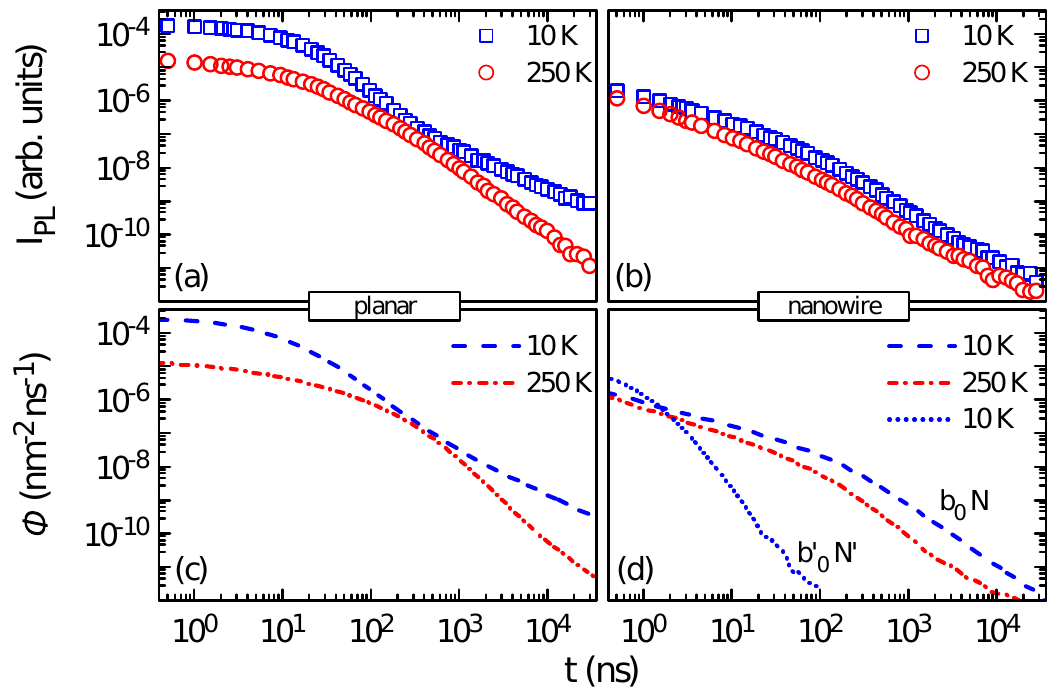}}
\end{figure*}

Figures \hyperref[fig:pl-lifetimes]{\ref*{fig:pl-lifetimes}(b)} and \hyperref[fig:pl-lifetimes]{\ref*{fig:pl-lifetimes}(c)} show the temperature dependence of the radiative and nonradiative lifetimes determined as described above for the planar and the nanowire sample, respectively. The PL lifetimes and the values of the IQEs at 10 and 250\,K are also compiled in Table~\ref{tab:lifetimes_iqe}. Regardless of the absolute values, we obtain a qualitative understanding of the decay processes by examining the temperature dependence of $\tau_r$. For both samples, $\tau_r$ is constant up to a certain temperature and then smoothly approaches a linear increase as indicated by the dashed lines. Hence, emission takes place from zero-dimensional (0D) localized states at low temperatures and approaches the behavior expected for a two-dimensional (2D) system (i.\,e., a QW) at higher temperatures \cite{Waltereit2001}. The transition from 0D to 2D occurs at a higher temperature and is more gradual for the nanowire sample, indicating that carriers in the QDs experience stronger localization than those in the planar QWs.

Regarding the absolute values of the lifetimes, we first see that the radiative lifetimes measured for the nanowire sample are significantly longer than those of the planar sample. This finding is consistent with the stronger polarization fields in the QDs expected from the higher In content, but also with stronger localization. Second, we see that $\tau_{nr}$ becomes shorter than $\tau_r$ at about 200\,K for the planar sample, while $\tau_{nr}$ is always drastically shorter than $\tau_r$ for the nanowire sample. Thus, the recombination in this sample is dominated by nonradiative recombination over the entire temperature range. 


\subsection{Analysis and simulation of the PL decay over the full time range}\label{sec:powerlaw}

The analysis of the initial PL decay discussed in Sec.~\ref{sec:trpl-ini} has provided useful information, but did not yield any insights as to the nonexponential nature of the PL decay. In early studies, this nonexponential decay invariably observed for (In,Ga)N/GaN(0001) QWs was proposed to be represented by a stretched exponential \cite{Pophristic1998}. This observation was attributed to strong compositional fluctuations in the ternary alloy (In,Ga)N, creating In-rich regions resembling ``quantum dots'' that confine excitons with different energies and, consequently, lifetimes \cite{Bartel2004}.

Alternatively, \citet{Morel2003} attributed the nonexponential decay to the recombination of individually localized electrons and holes that are separated spatially in the 2D QW plane. Their model was inspired by the seminal work of \citet{Thomas1965}, which  describes the kinetics of radiative recombination of electrons and holes bound to randomly distributed donor and acceptor pairs (DAPs) in a bulk crystal and is therefore known as the 2D-DAP model.

More recently, \citet{Brosseau2010} recorded PL transients of (In,Ga)N/GaN(0001) QWs over six decades in intensity. These measurements demonstrated that only the initial decay follows an exponential or stretched exponential dependence. For longer times, the decay was observed to deviate from this dependence and to asymptotically enter a power law. The authors analyzed their data by a phenomenological model based on the coexistence of a radiative state and a metastable charge-separated state \cite{Sher2008}. This model also described the experimentally observed asymptotic slowdown of the decay, which cannot be accounted for by the model of \citet{Morel2003}. \citet{Cardin2013} extended this study by investigating the PL decay kinetics of (In,Ga)N/GaN$(000\bar{1})$ nanowire heterostructures at room temperature and observed a power law decay for these structures as well.

A power law decay of the PL intensity is by no means restricted to (In,Ga)N, but is actually observed for the majority of solids \cite{Jonscher_JPC_1984}. The unifying characteristics of these materials is topological disorder as observed, for example, in solids with randomly situated traps such as found in various amorphous semiconductors. A wealth of studies is available on this subject, and its understanding is in fact much more mature than in the case of (In,Ga)N \cite{Tsang1979, Noolandi1980, Hong1981, Doktorov1982, Gourdon1989, Chen1992, Huntley2006, Seki2006}. For example, the impact of carrier diffusion on the PL decay has been studied in great detail already decades ago \cite{Hong1981}. However, nonradiative processes have not been taken into consideration in these previous studies, but are obviously essential for a full understanding of the PL decay in (In,Ga)N.

Figures~\hyperref[fig:exp-simu]{\ref*{fig:exp-simu}(a)} and \hyperref[fig:exp-simu]{\ref*{fig:exp-simu}(b)} show exemplary PL transients of our samples over the full time range of 30\,\textmu s. The double-logarithmic representation of the transients demonstrates that the initial decay analyzed in Fig.~\ref{fig:pl-lifetimes} amounts only to a fraction of the entire decay. Furthermore, this representation facilitates the direct identification of a power law decay.

Several important observations can be made from the experimental PL transients shown in Figs.~\hyperref[fig:exp-simu]{\ref*{fig:exp-simu}(a)} and \hyperref[fig:exp-simu]{\ref*{fig:exp-simu}(b)}. First, we observe a PL decay that obeys a power law asymptotically for both samples and independent of temperature. At 10\,K, the pronounced slowdown of the decay of the planar sample after about 300\,ns closely resembles the asymptotic behavior reported by \citet{Brosseau2010}.
Second, we observe a reduction of the temporally integrated PL intensity (area under the PL transients) with increasing temperature for both samples, reflecting the presence of nonradiative recombination at elevated temperatures. Despite this fact, the decay does not become exponential, but still obeys a power law for both samples at elevated temperatures. For the planar sample, the decay accelerates with temperature (the exponent of the $t^\beta$ asymptote increases from $\beta=-1.1$ at 10\,K to $\beta=-1.9$ at 250\,K), while the shape of the PL decay of the nanowire sample hardly changes at all between 10 and 250\,K.

\begin{figure}[b]
\includegraphics[width=0.73\columnwidth]{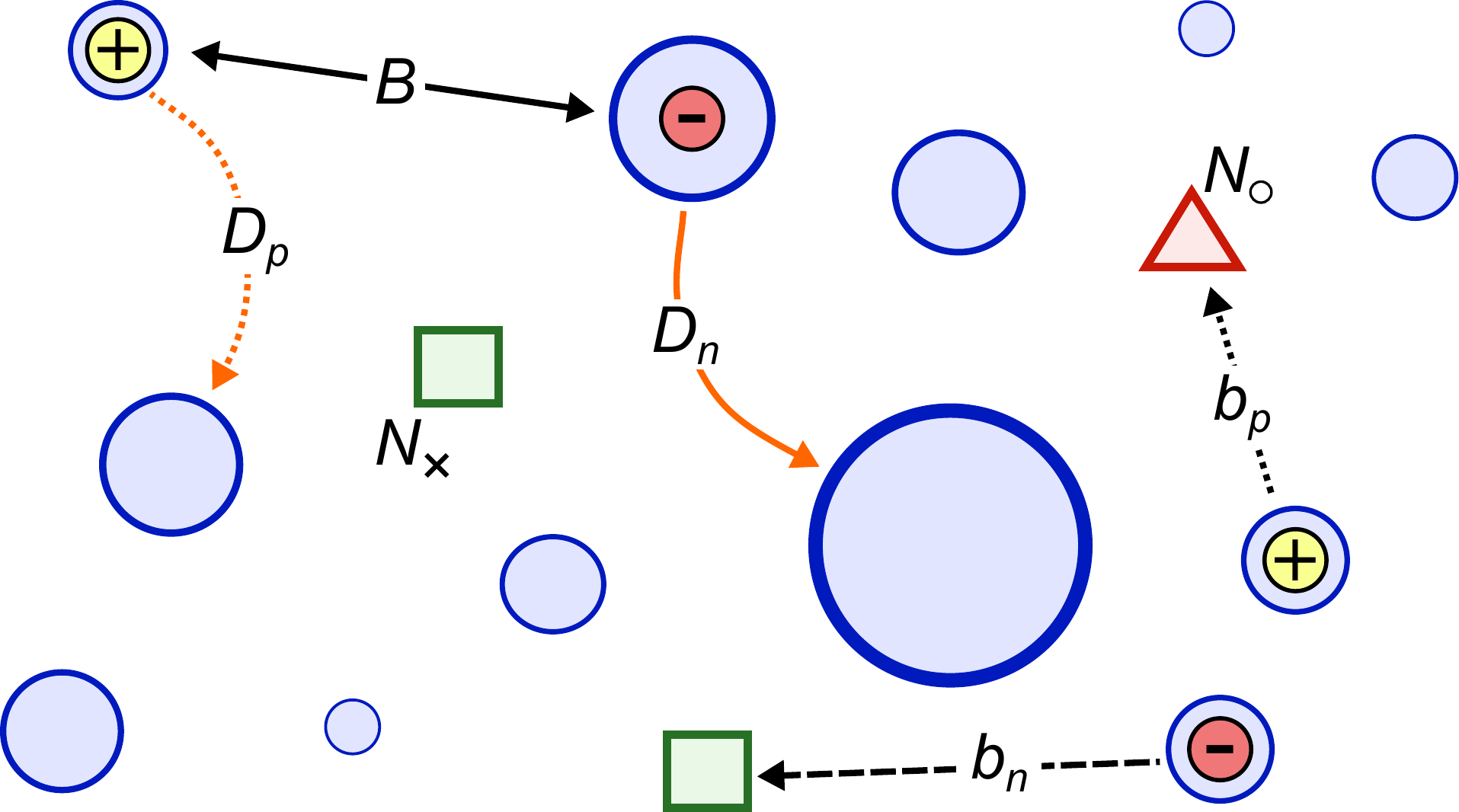}
\caption{Schematic representation of the model for simulating the PL transients of (In,Ga)N/GaN\{0001\} QWs. The potential landscape of the random alloy (In,Ga)N is assumed to create localization sites ($\bigcirc$) for electrons ($\ominus$) and holes ($\oplus$) with varying energy depth (represented by the diameter). In addition, nonradiative recombination centers ($\square$ and $\triangle$) exist. Initially, electrons and holes are randomly distributed. Radiative recombination occurs via tunneling with a coefficient $B$. The recombination centers capture electrons and holes by tunneling with coefficients $b_n$ and $b_p$, respectively. Diffusion (curved arrows) of electrons and holes allows them to migrate in the potential landscape with coefficients $D_n$ and $D_p$, respectively. All coefficients are random functions of location $\mathbf{r}$ and, for recombination events, of spatial distance $\mathbf{x}$.}
\label{fig:model}
\end{figure}

For a quantitative understanding of these transients, we consider the processes schematically depicted in Fig.~\ref{fig:model}. First of all, we assume that the power law decay is fundamentally related to the recombination of spatially separated electrons and holes. We also assume that (In,Ga)N constitutes a perfect random alloy \cite{Galtrey2007,Schulz2012} whose inherent compositional fluctuations are sufficient to localize charge carriers \cite{Brandt2006,Watson-Parris2011,Schulz2015c}. The potential landscape used in the following is constituted by randomly situated localization sites with a randomly varying energy depth, as visualized in Fig.~\ref{fig:model} by circles with different diameters. Due to the presence of nonradiative recombination, we include Shockley-Read-Hall recombination centers with a density $N$ that are also assumed to be randomly distributed. These centers are either in state $N_{\times}$ (represented by squares) and interact with electrons or in state $N_{\medcirc}=N-N_{\times}$ (represented by triangles) and interact with holes.

Initially, electrons (minus) and holes (plus) are distributed randomly at localization sites (cf.\ Fig.~\ref{fig:model}). The charge carriers can recombine radiatively via tunneling over distances $|\mathbf{x}|$ with a rate coefficient $B(|\mathbf{x}|,\mathbf{r})=B_0\exp[-|\mathbf{x}|/a(\mathbf{r})]$. Likewise, electrons and holes can be captured by recombination centers with rate coefficients  $b_n(|\mathbf{x}|,\mathbf{r})=b_{n0}\exp[-|\mathbf{x}|/a(\mathbf{r})]$ and $b_p(|\mathbf{x}|,\mathbf{r})=b_{p0}\exp[-|\mathbf{x}|/a(\mathbf{r})]$, respectively. The tunneling parameter $a(\mathbf{r})$ depends on the localization energy and is thus a function of the spatial location $\mathbf{r}=(x_i,x_j)$. Finally, electrons and holes are given the possibility to diffuse within this potential landscape with diffusivities $D_n(\mathbf{r})$ and $D_p(\mathbf{r})$, respectively. The diffusivity is influenced by the potential depth of the localization site and thus explicitly depends on $\mathbf{r}$.

\begin{table*}
\caption{Parameters used for the Monte Carlo simulation of the temperature-dependent PL transients of the planar and the nanowire sample shown in Figs.~\hyperref[fig:exp-simu]{\ref*{fig:exp-simu}(c)} and \hyperref[fig:exp-simu]{\ref*{fig:exp-simu}(d)}, respectively.}
 \label{tab:simu_para}
\begin{ruledtabular}
\begin{tabular}{c c c c c c c}
sample & $T$ (K) &$a$ (nm) & $D$ ($\text{nm}^2\,\text{ns}^{-1}$) & $B_0$ (ns$^{-1}$) & $b_0$ (ns$^{-1}$) & $N$ (nm$^{-2})$\\
\hline
planar & 10 & $20\pm 0.1$ & $0$ & $2\times 10^{-2}$ & $0$ & $3.1 \times 10^{-4}$\\
planar & 250 & $8\pm 0.1$ & $2.00$ & $1\times 10^{-3}$ & $1.0\times 10^{0}$ &  $3.1 \times 10^{-4}$\\
nanowire & 10 & $5\pm 2$ & $0$ & $1\times 10^{-2}$ & $2.8\times 10^{2}$ &  $9.4 \times 10^{-4}$\\
nanowire & 250 & $5\pm 1$ & $0.05$ & $1\times 10^{-2}$ & $4.0\times 10^{2}$ &  $9.4 \times 10^{-4}$
\end{tabular}
\end{ruledtabular}
\end{table*}

The three processes considered above (radiative recombination, Shockley-Read-Hall recombination, and carrier diffusion) are those considered in the classical diffusion-recombination equations of semiconductor physics \cite{Ahrenkiel1993}. Here, we generalize these equations by taking into account the stochastic nature of the recombination coefficients (i.\,e., the random dependence on position or distance of the respective process). We thus arrive at the following coupled system of integro-differential equations~(\ref{eq:3})--(\ref{eq:5}) for electrons $n$, holes $p$, and nonradiative centers $N$:

\begin{widetext}
\begin{align}
\label{eq:3}
\frac{\partial n(\mathbf{r}; t)}{\partial t} &= D_n(\mathbf{r})\Delta n(\mathbf{r}; t) -n(\mathbf{r}; t) \int B(|\mathbf{x}|,\mathbf{r})p(\mathbf{r} + \mathbf{x}; t)d\mathbf{x} -n(\mathbf{r}; t) \int b_n(|\mathbf{x}|,\mathbf{r})N_{\times}(\mathbf{r} + \mathbf{x}; t)d\mathbf{x}\\
\label{eq:4}
\frac{\partial p(\mathbf{r}; t)}{\partial t} &= D_p(\mathbf{r})\Delta p(\mathbf{r}; t) - p(\mathbf{r}; t) \int B(|\mathbf{x}|,\mathbf{r})n(\mathbf{r} + \mathbf{x}; t)d\mathbf{x} -p(\mathbf{r}; t) \int b_p(|\mathbf{x}|,\mathbf{r})[N(\mathbf{r}+\mathbf{x})-N_{\times}(\mathbf{r} + \mathbf{x}; t)]\,d\mathbf{x}\\
\label{eq:5}
\frac{\partial N_{\times}(\mathbf{r}; t)}{\partial t} &= - n(\mathbf{r}; t) \int b_n(|\mathbf{x}|,\mathbf{r})N_{\times}(\mathbf{r} + \mathbf{x}; t)d\mathbf{x} + p(\mathbf{r}; t) \int b_p(|\mathbf{x}|,\mathbf{r})[N(\mathbf{r}+\mathbf{x})-N_{\times}(\mathbf{r} + \mathbf{x}; t)]\,d\mathbf{x}
\end{align}
\end{widetext}

\noindent
The first term in Eqs.~(\ref{eq:3}) and (\ref{eq:4}) represents the diffusion of electrons and holes, respectively, while the second and third term represent the radiative and the nonradiative recombination of the respective type of charge carrier (either electrons or holes). The temporal evolution of the recombination centers in state $N_{\times}$ [Eq.~(\ref{eq:5})] is determined by the capture of an electron (first term) and the capture of a hole (second term) resulting in the nonradiative annihilation of both particles.

\begin{figure*}
\includegraphics[width=13cm]{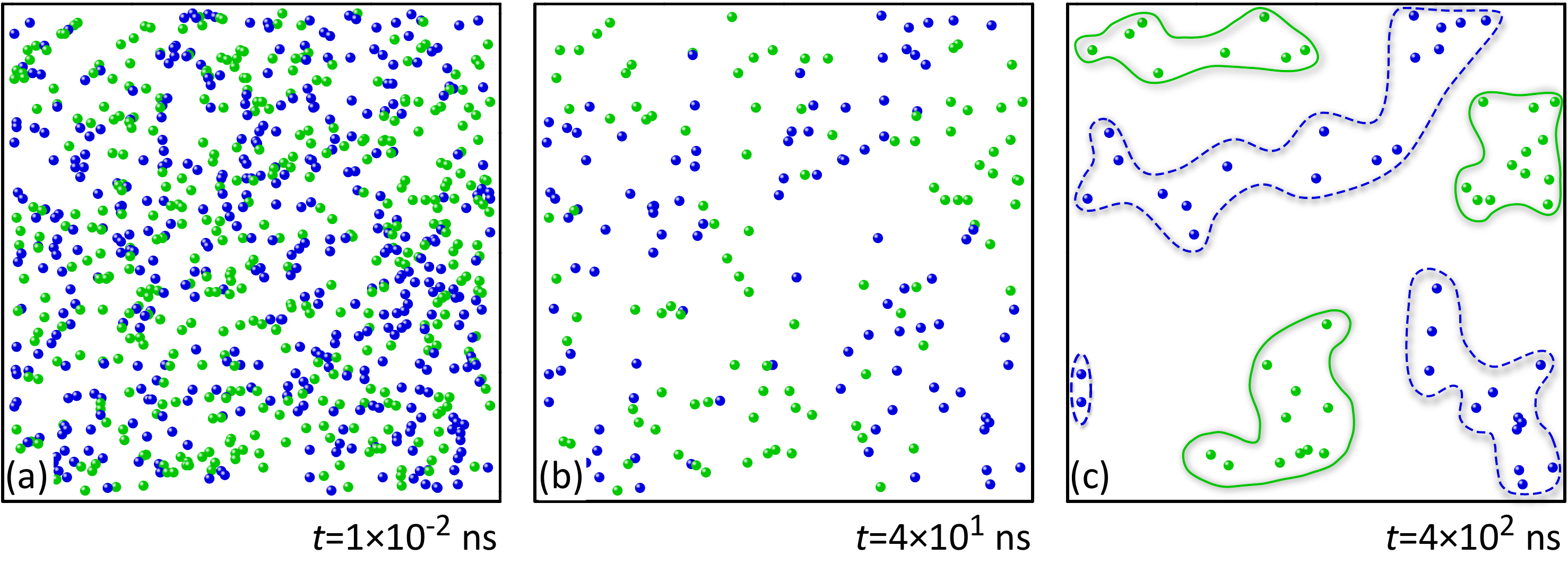}
\caption{Snapshots of the distribution of electrons (green) and holes (blue) during the simulated decay of the PL intensity of the planar sample at 10\,K. The snapshots are $400 \times 400$\, nm$^2$ in size and were taken (a) right after excitation, (b) at 40\,ns, and (c) at 400\,ns, after which the decay slows down significantly. The emerging clusters of electrons and holes are encircled by solid and dashed lines, respectively. For clarity, we do not display the (inactive) nonradiative recombination centers, and we show electrons and holes as spheres of uniform size.}
\label{fig:snapshot}
\end{figure*}

To solve these Smoluchowski-type equations \cite{Smoluchowski1916} numerically, we employ the Monte Carlo algorithm developed and described in detail in Ref.~\cite{Sabelfeld2015}.
Since the out-of-plane separation of electrons and holes is limited by the width of the QWs, which is on the order of the exciton Bohr radius, the distance between the recombination partners will be governed by their lateral (in-plane) separation, particularly when the decay of the PL approaches the power law asymptotically. We can thus simplify the problem considerably by reducing it to two dimensions.

For the actual simulations, the average density of localization sites is set to a value of 1\,nm$^{-2} $, which translates into one localization site every three by three unit meshes in the wurtzite lattice of (In,Ga)N. This density is roughly equivalent to the density of localized states obtained by means of atomistic tight-binding calculations for In$_{0.25}$Ga$_{0.75}$N \cite{Schulz2015c}. The charge carriers, with a density of $n = p = 2.5\times 10^{-3}$\,nm$^{-2}$ corresponding to the excitation density employed for the time-resolved experiments, are randomly distributed at localization sites at $t=0$. We do not allow one site to be occupied by more than one particle. Furthermore, in accordance to recent theoretical results \cite{Schulz2015}, we assume that the holes are localized at all temperatures ($D_p = 0$), but allow for a finite diffusivity of electrons ($D_n = D$) at elevated temperatures. The diffusion of electrons is implemented into the Monte Carlo algorithm by a hopping process between the different localization sites as explained in Ref.~\cite{Sabelfeld2015}. For simplicity, the nonradiative recombination coefficients for electrons and holes are set to be equal ($b_{n0} = b_{p0} = b_0$), and we assume that $N=N_{\times}$ at $t=0$. Additionally, charge carriers, once captured by a nonradiative center, will not be released again.

We have performed over $30{,}000$ simulations covering a large  parameter space to identify the parameter ranges reproducing the experimental transients (for a few selected examples, see Ref.~\cite{Sabelfeld2015}). Figures \hyperref[fig:exp-simu]{\ref*{fig:exp-simu}(c)} and \hyperref[fig:exp-simu]{\ref*{fig:exp-simu}(d)} show simulated transients in comparison to the experimental ones displayed in Figs.~\hyperref[fig:exp-simu]{\ref*{fig:exp-simu}(a)} and \hyperref[fig:exp-simu]{\ref*{fig:exp-simu}(b)}, respectively. The input parameters used for these simulations are listed in Table~\ref{tab:simu_para}. For a direct comparison with the simulated photon flux $\phi$, the experimentally measured, spectrally integrated PL intensity $I_\text {PL}$ is scaled by the same factor for both samples.

Let us first discuss the results for each sample separately. For the planar sample at 10\,K, the experimental transient with its pronounced slowdown at 400\,ns is only reproduced adequately when setting both the diffusivity and the nonradiative rate to zero, i.\,e., the simulated transient corresponds to the purely radiative recombination of localized electrons and holes. The peculiar slowdown of the decay is indeed a fingerprint for an IQE of unity, in agreement with our assumption in Secs.~\ref{sec:basic} and~\ref{sec:trpl-ini}. At 250\,K, the simultaneous loss in intensity and the acceleration of the decay are obtained by decreasing the radiative rate and enabling both electron diffusion and nonradiative recombination (cf.\ Table \ref{tab:simu_para}). The temperature dependence for the radiative and nonradiative processes are consistent with the respective lifetimes depicted in  Fig.~\hyperref[fig:pl-lifetimes]{\ref*{fig:pl-lifetimes}(b)}. Note that the nonradiative process conserves the power law decay, but eliminates the slowdown at 400\,ns as also observed experimentally.

Regardless of temperature, the experimental transients of the nanowire sample are characterized by a complete absence of an initial exponential phase [cf. Fig.~\ref{fig:pl-lifetimes}(a)]. The shape and intensity of these transients can only be reproduced by a dominant nonradiative process that depends very strongly on the carrier density. We obtain this strong density dependence by assuming capture coefficients that are more than two orders of magnitude larger than those of the planar sample, while the density of the centers is similar (this finding will be discussed in more detail below). The almost rigid downshift of the transient recorded at 250\,K in Fig.~\ref{fig:exp-simu}(b) is obtained by a further increase of the capture rate, while the radiative recombination rate does not change at all. Note that the radiative rate is smaller than that observed for the planar sample, in agreement with the results in Sec.~\ref{sec:trpl-ini}. Furthermore, diffusion is almost absent even at 250\,K, which directly reflects that carrier localization is significantly stronger in the (In,Ga)N QDs as compared to planar (In,Ga)N QWs. This result confirms the conclusions drawn from the results presented in Figs.~\hyperref[fig:cw-pl]{\ref*{fig:cw-pl}(a)} and \hyperref[fig:pl-lifetimes]{\ref*{fig:pl-lifetimes}(c)} and is also in agreement with the study of \citet{Laehnemann2014} on the same  nanowire sample.


Finally, we discuss four important issues in connection with these simulations: (i) the physical origin of the slowdown in the experimental and simulated PL transients of the planar sample shown in Figs.~\hyperref[fig:exp-simu]{\ref*{fig:exp-simu}(a)} and \hyperref[fig:exp-simu]{\ref*{fig:exp-simu}(c)}, (ii) the different role of $b_0$ and $N$ for the total nonradiative rate, (iii) the definition of a minority carrier diffusion length for (In,Ga)N, and (iv) the impact of carrier diffusion on the IQE.

(i) Figure \ref{fig:snapshot} shows snapshots of the spatial distribution of electrons and holes during their radiative recombination resulting in the simulated PL transient at 10\,K as shown in Fig.~\hyperref[fig:exp-simu]{\ref*{fig:exp-simu}(c)} (a complete video can be found in the supporting information).  Immediately after their excitation, electrons and holes are distributed randomly [Fig.~\ref{fig:snapshot}(a)]. Obviously, the electrons and holes most likely to recombine first are those with the least spatial separation. Hence, electrons in close vicinity to holes will disappear and vice versa. As a consequence, clusters of each individual species should emerge from the initial random distribution as indeed seen in the snapshot shown in Fig.~\ref{fig:snapshot}(b). These clusters become entirely spatially separated with continuing recombination [Fig.~\ref{fig:snapshot}(c)], and it is at this point where subsequent recombination slows down. This segregation phenomenon only occurs in the absence of nonradiative recombination, since nonradiative centers, being also randomly distributed, exist within electron as well as hole clusters, and diffusion, which constantly redistributes the electrons and constitutes the rate-limiting step determining the speed of recombination.

(ii) For classical Shockley-Read-Hall recombination, the steady-state recombination rate is proportional to the product $b N$, and one cannot distinguish an increase in the density of nonradiative recombination centers from an increase of the capture coefficient \cite{Brandt1996}. In the present case, however, the impact of $b=b_n=b_p$ and $N$ is different and can be distinguished. This fact is illustrated by the simulated transient labeled $b'_0 N'$ in Fig.~\hyperref[fig:exp-simu]{\ref*{fig:exp-simu}(d)}, for which we assumed the same value for the capture coefficient as for the planar sample ($b_0=1\times 10^0$\,ns$^{-1}$), but increased $N$ to the value required ($2 \times 10^{-2}$\,nm$^{-2}$) to keep the product $b N$ constant (taking into account that $|\mathbf{x}|$  equals asymptotically $ 1/2 \sqrt{N}$). Evidently, the two transients computed with the same value of $b N$ are drastically different. The origin of this different impact of $b$ and $N$ lies in the fact that the capture rate $b$ depends exponentially on the density of the centers as well as of the carriers. A higher prefactor for the nonradiative capture coefficient is thus not equivalent to a higher density of nonradiative centers.

(iii) The precise shape of the power law decay depends sensitively on the carrier diffusivity as already shown in Ref.~\cite{Sabelfeld2015}. Analyzing experimental transients by our recombination-diffusion model allows us to assess diffusion processes taking place on a nanometer scale without requiring any spatial resolution. In fact, even diffusivities as small as $10^{-5}\,\text{cm}^2\,\text{s}^{-1}$ can be detected, a value too small to be resolved by most other techniques. Note that this diffusivity cannot be translated into a diffusion length in the conventional sense since the diffusing species do not have a unique lifetime. Instead, both the lifetime of particles and their hopping distance within this lifetime are instantaneous quantities that vary over orders of magnitude with time. In the frame of our Monte Carlo simulation, we keep track of each individual particle including all of their elementary hops and recombination events. Averaging over all particles for a time up to 100\,\textmu s, we obtain a hopping distance of 13\,nm for the planar sample at 250\,K. Note, however, that this hopping distance is not distributed normally. A few particles do not hop even once before they recombine, but some migrate distances of several 100~nm.

(iv) Intuitively, one expects a monotonous increase of the IQE with either increasing excitation density, decreasing density of recombination centers, or smaller capture coefficients. Our model actually confirms this expectation (not shown here). However, the impact of diffusion on the IQE is not as straightforward. In general, diffusion processes accelerate the decay of the PL intensity. For the parameters used for simulating the transient for the planar sample at 250\,K [Fig.~\hyperref[fig:exp-simu]{\ref*{fig:exp-simu}(c)}], the diffusion of electrons is found to \emph{enhance} the IQE since it favors radiative recombination of neighboring electrons and holes in the initial phase of the PL decay over a capture by the nonradiative centers at a later stage. The opposite situation occurs for a sufficiently high density of nonradiative recombination centers, for which diffusion enhances nonradiative recombination over radiative one and thus results in a \emph{decline} of the IQE. 

\section{SUMMARY AND CONCLUSION}\label{sec:summary}
Our comparison of planar (In,Ga)N/GaN(0001) QWs and (In,Ga)N/GaN$(000\bar{1})$ QDs in nanowires has resulted in several important insights. First of all, we have shown that the recombination dynamics in the latter structures is  characterized by both strong carrier localization and a highly efficient nonradiative decay channel. Using the ratio of the PL intensities at high and low temperatures as a measure for the IQE may result in grossly overestimated values, and may thus be entirely misleading. The actual IQE of the (In,Ga)N QDs is low ($\approx 0.5$\%) even at 10\,K, but decreases only slightly to about 0.2\% at 250\,K thanks to the fact that localization prevails up to high temperatures. These values are consistent with the low EQEs reported for (In,Ga)N/GaN$(000\bar{1})$-based nanowire LEDs \cite{Janjua2016,Zhao2016,Zhao2016b}. They are also consistent with the peak EQE of 0.055\% measured for LEDs that we have fabricated from nanowire ensembles comparable to the one investigated in the present work \cite{Musolino2015a}.

An alternative and more reliable method to quantitatively investigate the IQE as well as carrier diffusion is the analysis of the PL transient recorded over a time interval sufficient to yield at least six decades of intensity. For both (In,Ga)N/GaN(0001) QWs and (In,Ga)N/GaN$(000\bar{1})$ QDs, a power law decay is observed, reflecting that recombination occurs between individual electrons and holes with varying spatial separation. The PL transient of the QWs at 10\,K exhibits a characteristic slowdown after about 400\,ns, which we have found to be a fingerprint of purely radiative recombination. Nonradiative recombination and carrier diffusion set in at 250\,K and eliminate this slowdown, but preserve the powerlaw decay. Even the much faster nonradiative recombination in the QDs does not result in an overall faster decay. A slow decay is thus not a reliable indicator for a high IQE. However, the shape of the transients observed for the QDs is quite different from those of the QWs and can be reproduced in simulations only when assuming recombination centers with very high capture coefficients. It seems likely that these centers are identical to those hypothesized to be responsible for the complete lack of an (In,Ga)N-related band in the PL spectra of planar (In,Ga)N/GaN$(000\bar{1})$ QWs grown by MBE \cite{Fernandez2016}. The strong localization effects observed for QDs in nanowires \cite{Laehnemann2014} may help a small fraction of the carrier population to evade nonradiative annihilation and may thus prevent the total dominance of nonradiative processes observed for homogeneous (In,Ga)N/GaN$(000\bar{1})$ QWs.

Our study has, however, not only provided insights into the eligibility of N-polar (In,Ga)N/GaN nanowires for use in future light emitters, but has also contributed to the understanding of materials used presently for commercial devices. Just as in (In,Ga)N/GaN$(000\bar{1})$ QDs, recombination in (In,Ga)N/GaN(0001) QWs occurs predominantly between individually localized, spatially separated electrons and holes. The recombination rates, whether radiative or nonradiative, thus depend strongly (in fact exponentially) on the carrier density. This dependence manifests itself in the highly nonexponential nature of the PL decay with its power law asymptotics. Obviously, analyzing the luminous efficiency of a  material with this characteristics by means of a model with constant recombination coefficients (i.\,e., the popular \emph{ABC} model \cite{Shen2007,Cho2013}) will lead to misleading results, as recently also pointed out by \citet{Badcock2016}. In addition, carrier diffusion has been found to occur at elevated temperatures and to affect the IQE, but is neglected in the \emph{ABC} model altogether. For understanding the origin of the droop of the EQE as well as the ``green gap'' in actual LEDs, models are required that go beyond the crude approximation offered by the \emph{ABC} model and properly describe the carrier dynamics in the material under consideration. The diffusion-reaction equations employed in the present work constitute a clear physical framework on which such a more general model could be based.
\section{ACKNOWLEDGMENT}
We are indebted to Philipp Drechsel (Osram Opto Semiconductors GmbH, Regensburg) for providing the (In,Ga)N/GaN(0001) multiple quantum well structure used in this study as a reference sample. We also thank Uwe Jahn for a critical reading of the manuscript. Financial support of this work within the European Union Framework Program FP7-NMP-2013-SMALL-7 under grant agreement No.\ 604416 (DEEPEN) and from the Russian Science Foundation under grant agreement No.\ 14-11-00083 is gratefully acknowledged.

%

\end{document}